\documentclass[aps,rmp,onecolumn,a4paper,superscriptaddress]{revtex4}
\pdfoutput=1
\usepackage{graphicx}
\usepackage{amsmath,amsfonts,amssymb}
\usepackage{pgf,tikz}
\usetikzlibrary{snakes,arrows,patterns}
\newcommand{\be}{\begin{equation}}
\newcommand{\ee}{\end{equation}}
\newcommand{\ba}{\begin{eqnarray}}
\newcommand{\ea}{\end{eqnarray}}

\begin{document}
\title{Density Functional Theory calculation on many-cores hybrid CPU-GPU architectures}

\author{Luigi Genovese}
\affiliation{European Synchrotron Radiation Facility, 6 rue Horowitz, BP 220, 38043 Grenoble France}
\email{luigi.genovese@esrf.fr, Tel.: +33 4 76 88 25 54}
\author{Matthieu Ospici}
\affiliation{Universit\'e Joseph Fourier - Laboratoire d'Informatique de Grenoble - INRIA, Grenoble, France}
\affiliation{Bull SAS, 1 rue de Provence, 38130 Echirolles, France}
\affiliation{Laboratoire de Simulation Atomistique (L\_Sim), SP2M/INAC/CEA, 17 Av. des Martyrs, 38054 Grenoble, France}
\author{Thierry Deutsch}
\affiliation{Laboratoire de Simulation Atomistique (L\_Sim), SP2M/INAC/CEA, 17 Av. des Martyrs, 38054 Grenoble, France}
\author{Jean-Fran\c{c}ois M\'ehaut}
\affiliation{Universit\'e Joseph Fourier - Laboratoire d'Informatique de Grenoble - INRIA, Grenoble, France}
\author{Alexey Neelov}
\affiliation{Institut f\"{u}r Physik, Universit\"{a}t Basel, Klingelbergstr. 82, 4056 Basel, Switzerland}
\affiliation{Institute for Computational Physics, Universit\"{a}t Stuttgart, Pfaffenwaldring 27, 70569 Stuttgart,Germany}
\author{Stefan Goedecker}
\affiliation{Institut f\"{u}r Physik, Universit\"{a}t Basel, Klingelbergstr. 82, 4056 Basel, Switzerland}


%

\begin{abstract}
The implementation of a full electronic structure calculation code on a hybrid parallel architecture with Graphic Processing Units (GPU) is presented.
The code which is on the basis of our implementation is a GNU-GPL code based on Daubechies wavelets.
It shows very good performances, systematic convergence properties and an excellent efficiency on parallel computers.
Our GPU-based acceleration fully preserves all these properties. In particular, the code is able to run on many cores which may or may not have a GPU associated.
It is thus able to run on parallel and massive parallel hybrid environment, also with a non-homogeneous ratio CPU/GPU.
With double precision calculations, we may achieve considerable speedup, between a factor of 20 for some operations and a factor of 6 for the whole DFT code.
\end{abstract}
\maketitle
\section{Introduction}
The Kohn-Sham (KS) formalism of the Density Functional Theory (DFT) approach is a well-established first-principles method for investigating properties of atomistic systems.
In the recent years, the increasing of the computational power of modern supercomputers has further stimulated the interest of the community for electronic structure calculations of systems with many electrons. Systems which were untractable only few years ago become now accessible with the advent of modern machines.
However, despite the approximate nature of the approach, the computational demand becomes huge already for systems with few hundreds atoms.
For the most distributed DFT codes, the number of computational operations scales cubically with respect to the number of atoms in the system.
As a result, the computational overhead for treating systems with large number of atoms now represents a serious limitation for the maximum size of the system considered.
A possible strategy to circumvent this problem can be provided by the development of linear scaling algorithms for electronic structure calculations, which have been widely developed in recent years. Such kind of approaches have a crossover point with the traditional cubic codes which is placed right around few hundreds atoms, and at present they represents one of the most promising strategies for \textit{ab initio} simulation of big systems that exhibit quantum mechanical behaviour.

In the past few years, the possibility of using Graphic Processing Units (GPU) for scientific calculations has raised a lot of interest.
A technology initially developed for home PC hardware has rapidly evolved in the direction of programmable parallel streaming processor.
The features of these devices, in particular the very low price performance ratio, together with the relatively low energy consumption, make them actractive platforms for intensive scientific computations.
A lot of scientific applications have been recently ported on GPU, including for example molecular dynamics \cite{mdGPU}, quantum Monte-Carlo \cite{QMCCPC}, Finite Element Methods \cite{parallelGPU}.
In the domain of electronic structure calculation, the calculation of the exchange-correlation term in a Gaussian based DFT code has been implemented on GPU \cite{GaussianGPU}, as well as the evaluation of the Coulomb potential \cite{TwoElecGPU}. Also the self consistent field calculation of the gaussian-based GAMESS code was ported on GPU \cite{GAMESSNVidia}.
Most of these implementation are performed on single precision calculation units, and with a single CPU core connected to a single GPU card.

In this paper we will present an implementation of a full DFT code which can run on massive parallel hybrid CPU-GPU clusters.
Our implementation is based on the achitecture of the most recent NVidia GPU cards (compute capability of type 1.3), which supports double precision floating point numbers.
The routines which define the GPU kernels are thus coded within the specification of the CUDA language \cite{CUDA}.
The DFT code which is on the basis of our implementation is a systematic code based on Daubechies wavelets\cite{daub}.
The latter is a systematic, orthogonal real-space basis set which presents optimal properties for expanding localised information.
The properties of this basis will turn out to be optimal for an extension on a GPU-accelerated environment.
This DFT code, named BigDFT \cite{BigDFT}, is a Free Software GNU-GPL code and is delivered either in a standalone version or integrated (in its pure CPU version) in the ABINIT~\cite{abinit} software package.

The paper is organised as follows: in the first section we will present the main features of the BigDFT code to see how the operators of the KS hamiltonian can be written in Daubechies wavelets basis.
We will then discuss the implementation of these operators in the GPU architectures, and inspect their performances separately from the complete DFT code.
Next, we will inspect the problem of the transfer of the data on the card, and we will show the performances of the full DFT code on parallel environments.

\section{Overview of the BigDFT code}
Let us briefly describe the main features of the DFT code on which our GPU developments are based.
The BigDFT code \cite{BigDFT}, is a free software (GNU-GPL) code issued from a three years STREP european project.
It is a pseudopotential DFT code, in which Kohn-Sham (KS) wavefunctions are expressed in Daubechies wavelets basis~\cite{daub}.
The latter is a multi-resolution basis, with compact-support, orthogonal functions.

The choice of the basis is of great importance in the computational operations which are performed by this code.
To explain this fact in more detail let us illustrate the main operations which must be performed in the context of a DFT calculation.

In the KS formulation of DFT, the KS wavefunctions $|\Psi_i\rangle$ are eigenfunctions of the KS Hamiltonian, with pseudopotential $V_\text{psp}$:
\be
\left(-\frac{1}{2} \nabla^2 + V_\text{KS}[\rho] + V_\text{psp} \right)|\Psi_i\rangle=\epsilon_i |\Psi_i\rangle\;.
\ee
The KS potential $V_\text{KS}[\rho]$ is a functional the electronic density of the system:
\be \label{density}
\rho(\mathbf r) = \sum_{i=1}^{\mathcal N_\text{orbitals}} n_\text{occ}^{(i)} \left|\Psi_i(\mathbf r)\right |^2\;,
\ee
where $n_\text{occ}^{(i)}$ is the occupation of orbital $i$.

The KS potential $V_\text{KS}[\rho] = V_H[\rho] + V_\text{xc}[\rho] + V_\text{ext}$ contains the Hartree potential $V_H$, solution of the Poisson's equation $\nabla^2 V_H = - 4 \pi \rho$, the exchange-correlation potential $V_\text{xc}$ and the external ionic potential $V_\text{ext}$ acting on the electrons. In BigDFT code the pseudopotential term $V_\text{psp}$ is of the form of norm-conserving GTH-HGH pseudopotentials~\cite{gth,hgh,krack}, which have a local and a nonlocal term, $V_\text{psp}=V_\text{local} + V_\text{nonlocal}$.
The KS hamiltonian can then be written as the action of three operators on the wavefunction:
\be
\left(-\frac{1}{2} \nabla^2 + V_L + V_\text{nonlocal} \right)|\Psi_i\rangle=\epsilon_i |\Psi_i\rangle\;,
\ee
where $V_L=V_H + V_\text{xc} + V_\text{ext} + V_\text{local}$ is a real-space based (local) potential, and $V_\text{nonlocal}$ comes from the pseudopotentials.

As usual in a KS DFT calculation, the application of the hamiltonian is a part of a self consistent cycle, needed for minimising the total energy.
In addition to the usual orthogonalization routine, in which scalar products $\langle \Psi_i | \Psi_j \rangle$ should be calculated, another operation which is performed on wavefunctions in BigDFT code is the preconditioning.
This is calculated by solving the Helmholtz equation
\begin{equation} \label{precon}
 \left( -\frac{1}{2} \nabla^2 - \epsilon_i \right) |\tilde{g}_i\rangle = |g_i\rangle \;,
\end{equation}
where $|g_i\rangle$ is the gradient of the total energy with respect to the wavefunction $|\Psi_i\rangle$, of energy $\epsilon_i$.
The preconditioned gradient $|\tilde{g}_i\rangle$ is found by solving Eq. \eqref{precon} by a preconditioned conjugate gradient method.

\subsection{Daubechies basis and wavelet transformation}
We describe in this section how the wavefunctions are expressed in the Daubechies basis. 
Though a more complete description can be found in \cite{BigDFT}, it is useful to revisit the principal points in view of a GPU implementation.

There are two fundamental functions in Daubechies family, the scaling function $\phi(x)$ and the wavelet $\psi(x)$. 
The full basis set can be obtained from all translations 
by a certain grid spacing $h$ of the scaling and wavelet functions centered at the origin.
Both functions are localized, with compact support. 
All the properties of these functions can be obtained from the relations 
\begin{align}
\label{refinement}
\phi(x) &= \sqrt 2\sum_{j=1-m}^{m} \text{\sl h}_j \: \phi(2 x -j) \\
\psi(x) &= \sqrt 2\sum_{j=1-m}^{m} \text{\sl g}_j \: \phi(2 x -j) \notag
\end{align}
which relate the basis functions on a grid with spacing $h$ and another 
one with spacing $h/2$. 
$\text{\sl h}_j$ and $\text{\sl g}_j=(-1)^j \text{\sl h}_{-j+1}$ are the elements of a filter that characterizes the wavelet family, and $m$ is the order of the family.
From Eq. \eqref{refinement}, every scaling function and wavelet on a coarse grid of spacing $h$ can be expressed as a linear combination of scaling functions at the finer grid level $h/2$. For this reason, wavelet functions complete the information which is lacking for refining the resolution level.
In our implementation we use only two resolution levels, namely one level of adaptivity. 
We can thus classify the coarse and the fine degrees of freedom by the information expanded by the scaling and the wavelet functions respectively.

For a three-dimensional description, the simplest basis set is obtained by a tensor product of one dimensional basis functions. For a two resolution level description, in each grid point the coarse degrees of freedom are expanded by a single three dimensional function $\phi^0_{i_1,i_2,i_3}(\mathbf r)$, while the fine degrees of freedom can be expressed by adding other seven basis functions, $\phi^{\nu}_{j_1,j_2,j_3}(\mathbf r)$, which include tensor products with one-dimensional wavelet functions.

A wavefunction $\Psi(\mathbf r)$ can thus be expanded in this basis:
\be \label{wavef0}
\Psi({\bf r}) = \sum_{i_1,i_2,i_3} c^0_{i_1,i_2,i_3} \phi^0_{i_1,i_2,i_3}(\mathbf r) + \sum_{j_1,j_2,j_3} \sum_{\nu=1}^7 c^\nu_{j_1,j_2,j_3} \phi^{\nu}_{j_1,j_2,j_3}(\mathbf r) 
\ee

The sum over $i_1$, $i_2$, $i_3$ runs over all the grid points contained in the low resolution region and the sum over $j_1$, $j_2$, $j_3$ over all the points contained in the (generally smaller) high resolution region. Such points belong to a uniform mesh of grid spacing $h$.
Each wavefunction is then associated to a set of coefficients $\{c^\mu_{j_1,j_2,j_3}\}$, $\mu=0,\cdots,7$.
The wavefunctions are stored in a compressed form where only the nonzero scaling function and wavelets coefficients are stored. The basis set being orthogonal, several operations such as scalar products among different orbitals and between orbitals and the projectors of the non-local pseudopotential can directly be done in this compressed form. 

The local hamiltonian operator can be applied in the pure fine scaling function representation, (a basis set which contains only scaling functions $\phi_{i'_1,i'_2,i'_3}$ centered on a finer grid of spacing $h'=h/2$):
\be
\label{wavefhigh}
\Psi({\bf r}) = \sum_{i'_1,i'_2,i'_3} s_{i'_1,i'_2,i'_3} \phi_{i'_1,i'_2,i'_3}(\mathbf r) \;.
\ee
The transformation between a mixed coarse scaling function/wavelet representation and a pure fine scaling function representation to  is done by the fast wavelet 
transformation~\cite{ppur} which is a three dimensional, separable convolution, that can be obtained from the filters $\text{\sl h}_j$ and $\text{\sl g}_j$ of Eq.\eqref{refinement}.

\subsection{The kinetic operator}
The matrix elements of the kinetic energy operator among the basis functions can be calculated analytically~\cite{beylkin,BigDFT}.
For the pure fine scaling function representation described in Eq. \eqref{wavefhigh}, the result of the application of the kinetic energy operator on this wavefunction, has the expansion coefficients $\hat{s}_{i'_1,i'_2,i'_3}$, which are related to the original coefficients $s_{i'_1,i'_2,i'_3}$ by a convolution
\begin{equation} \label{kiner}
 \hat{s}_{i'_1,i'_2,i'_3} =\frac{1}{2} \sum_{j'_1,j'_2,j'_3} K_{i'_1-j'_1,i'_2-j'_2,i'_3-j'_3} s_{j'_1,j'_2,j'_3} 
\end{equation}
where 
\begin{equation} \label{fprod}
K_{i_1,i_2,i_3}=T_{i_1}\delta_{i_2}\delta_{i_3} +\delta_{i_1} T_{i_2} \delta_{i_3}+ \delta_{i_1} \delta_{i_2} T_{i_3}\;,
\end{equation}
and $T_i$ are the filters of the one-dimensional second derivative in Daubechies scaling functions basis, which can be computed analytically.

\subsection{Application of the local potential}\label{magicfilters}
The potential $V_L$ is defined in real space, in particular on the points of the finer grid of spacing $h'$. 
The application of the local potential in Daubechies basis consists of the basis decomposition of the function product $V_L(\mathbf r) \Psi(\mathbf r)$.
As explained in \cite{magic,BigDFT}, the simple evaluation of this product in terms of the point values of the basis functions is not precise enough.
A better result may be achieved by performing a transformation to the wavefunction coefficients, which allows to calculate the values of the wavefunctions on the fine grid, via a smoothed version of the basis functions. This is the so-called ``magic filter'' transformation, which can be expressed as follows:
\be \label{magicfilter}
 \Psi(\mathbf r_{i'_1,i'_2,i'_3}) = \sum_{j'_1,j'_2,j'_3} \omega_{i'_1-j'_1}\omega_{i'_2-j'_2}\omega_{i'_3-j'_3} s_{j'_1,j'_2,j'_3} \;.
\ee
and allows to express with better accuracy the potential application. In other terms, the point values of a given wavefunction $|\Psi\rangle$ are expressed as if $\Psi(\mathbf r)$ would be the smoothest function which has the same Daubechies expansion coefficients of $|\Psi \rangle$. This procedure guarantees the highest precision ($\mathcal O(h^{16})$ in the potential energy) and can be computationally expressed by a three-dimensional separable convolution in terms of the filters $\omega_i$.
After application of the local potential (pointwise product), a transposed magic filter transformation can be applied to obtain Daubechies expansion coefficients of $V_L | \Psi \rangle$.

\subsection{Full local hamiltonian}
\label{locham}
The above described operations must be combined together for the application of the local hamiltonian $\left(-\frac{1}{2} \nabla^2 + V_L(\mathbf r)\right)$.
The order of the operation must be the following:
\begin{enumerate}
\item[1)] Decompression of the data on the whole simulation grid,
\item[2)] Wavelet transformation on the fine grid,
\item[3)] Local potential application:
\begin{enumerate}
\item[3a)] Magic filter transformation,
\item[3b)] Potential multiplication,
\item[3c)] Transposed magic filter transformation.
\end{enumerate}
\item[4)] Kinetic operator from the output of 2),
\item[5)] Sum with potential application,
\item[6)] Inverse wavelet transformation,
\item[7)] Data compression. 
\end{enumerate}
In the CPU version of BigDFT some of these operations are combined together, e.g. for the potential application 2) and 3a), and also  3c) and 6) are combined into a set of common filters, and the boundaries can be arranged such as to avoid data compression-decompression. However, in our GPU implementation we choose to separately apply these operations, which is simpler and accounts for better modularity. Moreover, different boundary conditions can be implemented immediately.

\subsection{Local density calculation}\label{locden}
The density of the electronic system is derived from the square of the point values of the wavefunctions, (see Eq. \eqref{density}). 
As described is section \ref{magicfilters}, a convenient way to express the point values of the wavefunctions is to apply the magic filter transformation to the wavefunctions expressed in Daubechies basis.
The operations needed for calculating the local density would then be identical to the points 1), 2) and 3a) of the list at section \ref{locham}, followed by an accumulation of the squares of the wavefunctions in the same array.

\subsection{Other operations}
The local potential $V_L$ can be obtained from the local density $\rho$ by solving the Poisson's equation and by calculating the exchange-correlation potential $V_{\text{xc}}[\rho]$.
These operations are performed via a Poisson Solver based on interpolating scaling functions \cite{lazy}, which is a basis set tightly connected with Daubechies functions,
optimal for electostatic problems, and which allows for mixed boundary conditions. A description of this Poisson solver can be found in the papers \cite{freePS, surfacePS}.

The complete hamiltonian contains also the nonlocal part of the pseudopotential which, thanks to the orthogonality of Daubechies wavelets, can directly be applied in the compressed form.
All the other operations can also be performed in the compressed form. In particular, overlap matrices needed for the orthogonality constraints, and their manipulations, are implemented by suitable applications of BLAS -LAPACK routines.

\subsection{Parallelization for homogeneous computing clusters (CPU code)}\label{parallelcpu}
The localisation and the orthogonality of the basis functions are of key importance for improving the performances of the code.
Indeed, thanks to their relatively  short filters, the convolutions can be efficiently optimised.
Two data distribution schemes are used for parallelising the code.
In the orbital distribution scheme (used for hamiltonian application, preconditioning), each processor works on one or  a few orbitals 
for which it holds all its scaling function and wavelet coefficients. The other operations are performed in the coefficient 
distribution scheme, where each processor holds a certain subset of the coefficients of all the orbitals.
For an homogeneous distribution between orbitals, switch back and forth between the orbital distribution scheme and the coefficient distribution scheme is done by the MPI global transposition routine MPI\_ALLTOALL. 
Further flexibility has been added by allowing each processor to store variable number of orbitals and/or wavefunction coefficients. 
This is particularly useful for an hybrid architecture, when only some of the cores can be in relation with some hardware accelerator (a GPU for example).
For this variable repartition case, the communication are performed with suitable calls to MPI\_ALLTOALLV routines.
For parallel computers where the cross sectional bandwidth~\cite{myoptimization} scales well with the number of processors this global transposition does not require a lot of CPU time.
The BigDFT code shows very good computational performances and an excellent efficiency (above 90\%) on parallel computers.

\section{Implementation of convolutions on GPU}
\subsection{Wavelet transformation and Magic Filters convolutions}\label{MFdescription}
As discussed above, the operations we are going to implement on GPU are essentially three-dimensional convolutions with \emph{short}, \emph{separable} filters.
In order to present the computational implementation, let us analyse in detail the magic filter transformation.
A three-dimensional array $s$ (input array) of dimension $n_1,n_2,n_3$ (the dimension of the grid of spacing $h'$) is transformed into the output array $\Psi_r$ given by
\begin{equation}\label{3dconv}
 {\Psi_r}(I_1,I_2,I_3) = \sum_{j_1,j_2,j_3=-L}^{U} \omega_{j_1} \omega_{j_2} \omega_{j_3} s(I_1-j_1,I_2-j_2,I_3-j_3)\;.
\end{equation}
With a lowercase index $i_p$ we indicate the elements of the input array, while with a capital index $I_p$ we indicate the indices after application of the magic filters $\omega_i$, which have extension from $-L$ to $U$. The filter dimension $U+L+1$ equals to the order of the Daubechies family $m$, which is 16 in our case.
In BigDFT, different boundary conditions (BC) can be applied at the border of the simulation region, 
which affects the values of the array $s$ in \eqref{3dconv} when the indices are outside bounds.
For our GPU implementation we use periodic BC in the three dimensions, i.e. data are wrapped periodically, tough the original CPU version of these convolutions also admits isolated and slab-like BC. However, our GPU implementation can be straightforwardly extended to these BC.

The most convenient way to calculate a three dimensional convolution of this kind is by combining one dimensional convolutions and array transpositions, as explained in \cite{stefanCPC}
In fact, the calculation \eqref{3dconv} can be cut in three steps:
\begin{enumerate}
\item[1)] $F_3(I_3,i_1,i_2) = \sum_{j} \omega_{j} s(i_1,i_2,I_3-j) $\qquad $\forall i_1,i_2$;
\item[2)] $F_2(I_2,I_3,i_1) = \sum_{j} \omega_{j} F_3(I_3,i_1,I_2-j) $\qquad $\forall I_3,i_1$;
\item[3)] $\Psi_r(I_1,I_2,I_3) = \sum_{j} \omega_{j} F_2(I_2,I_3,I_1-j) $\qquad $\forall I_2,I_3$.
\end{enumerate}
The final result is thus obtained by a successive application of the same operation:
\begin{equation}\label{1dconvtrans}
 F(I,a) = \sum_{j=-L}^{U} \omega_{j} G(a,I-j) \qquad \forall a=1,\cdots,N\,,\; I=1,\cdots,n\;.
\end{equation}
The lowest level routine which will be ported on GPU is then a set of $N$ separate, one dimensional (periodic), convolutions of arrays of size $n$.
The number $N$ equals $n_1 n_2$, $n_1 n_3$ and $n_2 n_3$ respectively, for each step of the three dimensional construction, while $n$ equals to the dimension which is going to be transformed.
The output of the first step is then taken as the input of the second, and so on.

\subsection{NVidia GPU Architecture}
A NVidia GPU is composed of a global memory and a set of multiprocessors. Each multiprocessor includes eight computing units (cores) and a "private" shared memory. To obtain optimal performance, it is necessary to store the data in the private memory of multiprocessors. The CUDA programming model \cite{CUDA} defines mechanisms allowing the data transfer in the shared memory. Fine-grained threads are the basic means of parallel execution in CUDA. Indeed, A CUDA program is mapped on a set ("grid") of execution blocks running on multiprocessors. A block contains several threads executed by the processors of a multiprocessor. The processors execute the threads in a synchronous way (data parallel threads).
In the following section, we will show the design and implementation of the Magic Filter convolution described in Sec.\ref{MFdescription} on NVIDIA GPUs. 

\subsection{Implementation details}
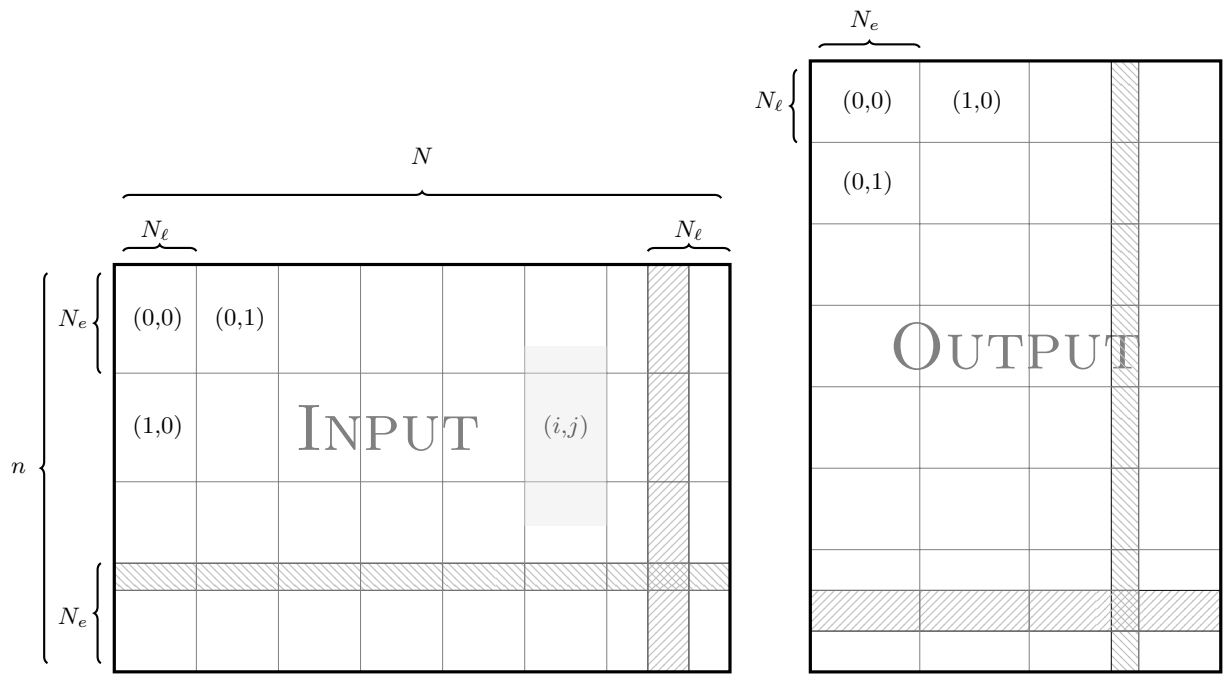
\begin{figure}
\begin{tikzpicture}[scale=1.8]
\node (no) at (0,0){};
\node (oo) at (0,3) {};
\node (ondat) at (4.5,3) {};
\node[opacity=0.5] (in) at (2,1.8) {\Huge \sc{Input}};
\draw[snake=brace,thick,raise snake=5pt] (oo) -- ++(0.6,0);
\draw[snake=brace,thick,raise snake=5pt] (3.9,3) -- ++(0.6,0);
\draw[snake=brace,thick,raise snake=5pt,mirror snake] (oo) -- ++(0,-0.8);
\draw[snake=brace,thick,raise snake=5pt] (no) -- ++(0,0.8);
\draw[snake=brace,thick,raise snake=25pt] (no) -- (oo);
\draw[snake=brace,thick,raise snake=25pt] (oo) -- (ondat);
\node (nlines) at (0.3,3.25) {$N_\ell$};
\node (nlines) at (4.2,3.25) {$N_\ell$};
\node (nlines) at (-0.3,0.4) {$N_e$};
\node (nlines) at (-0.3,2.6) {$N_e$};
\node (nlines) at (-0.7,1.5) {$n$};
\node (nlines) at (2.25,3.8) {$N$};
\draw[pattern=north west lines,very thin,pattern color=gray!50] (0,0.8) rectangle (4.5,0.6);
\draw[pattern=north east lines,very thin,pattern color=gray!50] (3.9,0) rectangle (4.2,3);
\node[text width=0.6cm,text centered] at (0.3,2.6) {(0,0)};
\node[text width=0.6cm,text centered] at (0.3,1.8) {(1,0)};
\node[text width=0.6cm,text centered] at (0.9,2.6) {(0,1)};
\node[text width=0.6cm,text centered] at (3.3,1.8) {($i$,$j$)};
\draw[help lines,xstep=0.6,ystep=0.8,yshift=0.6cm] (no) ++(0,0.6) grid +(4.2,2.4);
\draw[help lines,xstep=0.6,ystep=0.8] (no) grid +(4.2,0.8);
\draw[help lines,xstep=0.6,ystep=0.8,xshift=0.9cm,yshift=0.6cm] (no) ++(3.9,0.6) grid +(0.6,2.4);
\draw[help lines,xstep=0.6,ystep=0.8,xshift=0.9cm] (no) ++(3.9,0) grid +(0.6,0.8);
\draw[very thick] (no) rectangle (ondat);
\draw[opacity=0,dashed,fill=gray!20,fill opacity=0.4] (3,2.4) rectangle (3.6,1.08);
\end{tikzpicture}
\begin{tikzpicture}[scale=1.8,rotate=-90]
\node (no) at (0,0){};
\node (oo) at ++(0,3) {};
\node (ondat) at ++(4.5,3) {};
\node[opacity=0.5] (in) at (2.1,1.5) {\Huge \sc{Output}};
\draw[snake=brace,thick,raise snake=5pt,mirror snake] (no) -- ++(0.6,0);
\draw[snake=brace,thick,raise snake=5pt] (no) -- ++(0,0.8);
\node (nlines) at (0.3,-0.3) {$N_\ell$};
\node (nlines) at (-0.3,0.4) {$N_e$};
\draw[pattern=north west lines,very thin,pattern color=gray!50] (0,2.4) rectangle (4.5,2.2);
\draw[pattern=north east lines,very thin,pattern color=gray!50] (3.9,0) rectangle (4.2,3);
\node[text width=0.6cm,text centered] at (0.3,0.4) {(0,0)};
\node[text width=0.6cm,text centered] at (0.3,1.2) {(1,0)};
\node[text width=0.6cm,text centered] at (0.9,0.4) {(0,1)};
\draw[help lines,xstep=0.6,ystep=0.8] (no) ++(0,0) grid +(4.2,2.4);
\draw[help lines,xstep=0.6,ystep=0.8,yshift=0.6cm] (0,1.6) grid +(4.2,0.8);
\draw[help lines,xstep=0.6,ystep=0.8,xshift=0.9cm] (no) ++(3.9,0) grid +(0.6,2.4);
\draw[help lines,xstep=0.6,ystep=0.8,xshift=0.9cm] (no) ++(3.9,1.6) grid +(0.6,0.8);
\draw[very thick] (no) rectangle (ondat);
\end{tikzpicture}
\caption{Data distribution for 1D convolution+transposition on the GPU. Input data (left panel) are ordered along the $N$-axis, while output (right panel) is ordered in $n$-axis direction, see Eq. \eqref{1dconvtrans}. When executing GPU convolution kernel, each block ($i$,$j$) of the execution grid is associated to a set of $N_\ell$ ($N$-axis) times $N_e$ ($n$-axis) elements. The size of the data feed to each block is identical (such as to avoid block-dependent treatment), hence when $N$ and $n$ are not multiples of $N_\ell$ and $N_e$, some data treated by different blocks may overlap. This is indicated with the filled patterns in the figure. Behind the ($i$,$j$) label, in light gray, it is indicated the portion of data which should be copied to the shared memory for treating the data in the block. See Fig. \ref{shared} for a detail of that part.}
\label{transposition}
\end{figure}

From the GPU parallelism point of view, there is a set of N independent convolutions to be computed.
Each of the lines of $n$ elements to be transformed is split in chunks of size $N_e$.
Each multiprocessor of the graphic card computes a group of $N_\ell$ different chunks and parallellises the calculation on its computing units.
Figure \ref{transposition} shows the data distribution on the grid of blocks during the transposition.

\begin{figure}
\begin{center}
\begin{tikzpicture}[scale=1]
\node[opacity=0.5] at (2,8) {\huge\sc Data};
\node at (-0.5,-1.1) {$U$};
\node at (-0.5,6.75) {$L$};
\node at (2,6.5) {$N_\ell$};
\node at (-0.7,3) {$N_e$};
\draw[opacity=0,dashed,fill=gray!20,fill opacity=0.2] (0,-2.4) rectangle (4,7.5);
\draw[help lines,dashed,opacity=0.3,xstep=6,ystep=0.4] (0,-2.4) grid (4,7.5);
\draw[thick,dashed] (0,-2.2) rectangle (4,7.5);
\draw[very thick] (0,0) rectangle (4,6);
\draw[snake=brace,thick,raise snake=5pt,mirror snake] (0,7.5) -- ++(0,-1.5);
\draw[snake=brace,thick,raise snake=5pt,mirror snake] (0,0) -- ++(0,-2.2);
\draw[snake=brace,thick,raise snake=5pt,mirror snake] (4,6) -- ++(-4,0);
\draw[snake=brace,thick,raise snake=10pt] (0,0) -- ++(0,6);
\node at (2,5.4) [rectangle,opacity=0.4] {\texttt{warpSize/2}};
\end{tikzpicture}
\qquad
\begin{tikzpicture}[scale=1]
\node[opacity=0.5] at (2,8) {\huge\sc Copy};
\node[opacity=0.8] at (2,6.7) {\large (*,0)};
\node[opacity=0.8] at (2,5.1) {\large (*,1)};
\node[opacity=0.8] at (2,-1.8) {\large (*,$N_t-1$)};
\draw[opacity=0,dashed,fill=gray!20,fill opacity=0.2] (0,-2.4) rectangle (4,7.5);
\draw[help lines,dashed,opacity=0.3,xstep=6,ystep=0.4] (0,-2.4) grid (4,7.5);
\draw[help lines,xstep=6,ystep=1.6,yshift=-0.4cm] (0,1.5) grid (4,7.5);
\draw[help lines,xstep=6,ystep=1.2] (0,-2.4) grid (4,1.2);
\draw[thick,dashed] (0,-2.2) rectangle (4,7.5);
\draw[very thick] (0,0) rectangle (4,6);
\end{tikzpicture}
\qquad
\begin{tikzpicture}[scale=1]
\node[opacity=0.5] at (2,8) {\huge\sc Calc};
\node[opacity=0.8] at (2,5.4) {\large (*,0)};
\node[opacity=0.8] at (2,0.4) {\large (*,$N_t-1$)};
\draw[opacity=0,dashed,fill=gray!20,fill opacity=0.2] (0,-2.4) rectangle (4,7.5);
\draw[help lines,dashed,opacity=0.3,xstep=6,ystep=0.4] (0,-2.4) grid (4,7.5);
\draw[help lines,xstep=6,ystep=1.2] (0,2.2) grid ++(4,3.6);
\draw[help lines,xstep=6,ystep=0.8] (0,0) grid ++(4,2.2);
\draw[thick,dashed] (0,-2.2) rectangle (4,7.5);
\draw[very thick] (0,0) rectangle (4,6);
\end{tikzpicture}
\end{center}
\caption{Data arrangment in shared memory for a given block. The number of lines $N_\ell$ is chosen to be a divisor of the half-warp size. Data are then treated in units of \texttt{warpSize/2}. The thread index has $(\texttt{warpSize/2},N_t)$ elements, with $N_t \leq 16$ (left panel). Each group of threads $(*,i)$ of the half-warp $i$ treats a definite number of elements, either for copying the data (center panel) or for performing the calculation(right panel). This data arrangement ensures the avoiding of bank conflicts in the shared memory access. For calcuating the convolution, two buffers of sizes $N_\ell L$ and $N_\ell U$ must be created in shared memory. This figure reproduces the portion of the input data higlighted in gray in Fig. \ref{transposition}.}
\label{shared}
\end{figure}
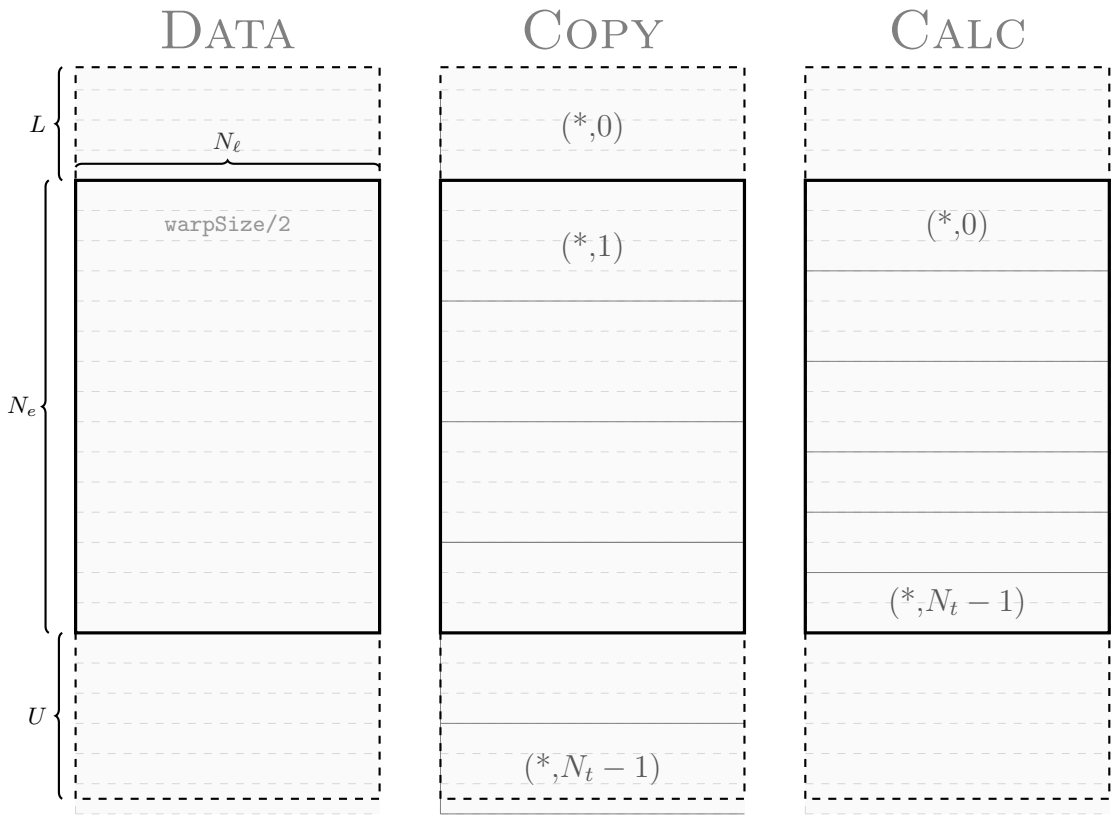

In order to get best performance with the GPU, its is strongly recommended to transfer data from the global to the shared memory of multiprocessors. 
The shared memory must contain buffers to store the data needed for the convulation computations. 
The desired boundary conditions (periodic in our case) is implemented in the shared memory during the data transfer.
Each thread computes the convolution for a subset of $N_e$ elements associated to the block.
This data distribution is illustrated in Fig. \ref{shared}.

\subsection{Kinetic convolution}
The GPU implementation described for the magic filter convolution can be applied also to the wavelet transformation, by suitably changing the values of the filters and their extensions.
This can be done thanks to the fact that also the latter is a three-dimensional separable convolution, i.e. it has the same formal expression as Eq.\eqref{3dconv}.
A little, but substantial, difference should be stressed for the kinetic operator application, defined in Eqs.\eqref{kiner} and \eqref{fprod}.
In this case the three dimensional filter is the sum of three different filters.
This implies that the kinetic filter operation must be cut differently from the other separable convolutions:
\begin{enumerate}
\item[1)] $K_3(I_3,i_1,i_2) = \sum_{j} T_{j} s(i_1,i_2,I_3-j) $\qquad $\forall i_1,i_2$;
\item[2)] $K_2(I_2,I_3,i_1) = \sum_{j} T_{j} s(i_1,I_2-j,I_3) + K_3(I_3,i_1,I_2) $\qquad $\forall I_3,i_1$;
\item[3)] $\hat{s}(I_1,I_2,I_3) = \sum_{j} T_{j} s(I_1-j,I_2,I_3) + K_2(I_2,I_3,I_1)$\qquad $\forall I_2,I_3$.
 \end{enumerate}
Also in this case, the three dimensional kinetic operator can be seen as a results of a successive application of the same operation:
\begin{equation}\label{kinetic1d}
K_p(I,a) = \sum_j T_{j} G_{p-1}(a,I-j) + K_{p-1}(a,I)\qquad \text{\emph{et}} \qquad G_p(I,a)=G_{p-1}(a,I)\qquad \forall a,I\;,
\end{equation}
In other terms, the one-dimensional kernel of the kinetic energy has two input arrays, $G_{p-1}$ and $K_{p-1}$, and returns two output arrays $K_p$ and the $G_p$, which is the transposition of $G_{p-1}$.
At the first step ($p=1$) we put $G_0=s$ and $K_0=0$. Eventually, for $p=3$, we have $K_3=\hat{s}$ and $G_3=s$.
One can put instead $K_0 \neq 0$, and in that way will be $K_3=\hat{s}+K_0$. This allows to unify step 4) and 5) in the list of section \ref{locham}, by putting $K_O$ equal to the output of step 3c).
This algorithm can be also used for the Helmholtz operator of the preconditioner, by putting $K_0=-\epsilon_i s$.

\section{Performance evaluation of GPU convolution routines}\label{perfmonocore}
To evaluate the performance of 1D convolution routines described in Eqs.\eqref{kinetic1d} and \eqref{1dconvtrans}, together with the analogous operation for the wavelet transformation, we are going to compare the execution times on CPU and GPU. We define the GPU speedup with the ratio between CPU and GPU execution times. For these evaluations, we used a computer with an Intel Xeon Processor X5472 (3 GHz) and a NVidia Tesla S1070 card. The CPU version of BigDFT is deeply optimized with optimal loop unrolling and compiler options. The GPU code is compiled with the Intel Fortran Compiler (10.1.011) and the most aggressive compiler options (\texttt{-O2 -xT}).
With these options the magic filter convolutions run at about 3.4 GFlops.
Since the GPU convolutions are called by an electronic structure code, all benchmarks are performed with the double precision floating point numbers proposed by the recent GPU. 

\begin{figure}
\begin{center}
\includegraphics[width=0.5\textwidth]{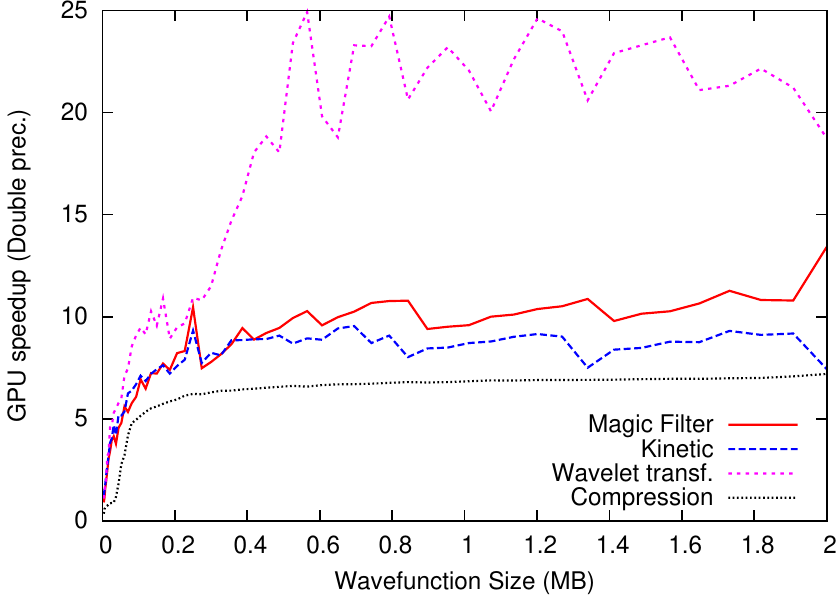}
\end{center}
\caption{Double precision speedup for the GPU version of the fundamental operations on the wavefunctions as a function of the single wavefunction size.}
\label{1dbench}
\end{figure}

The performance graphs for the three above mentioned convolutions, together with the compression-decompression operator, are indicated in Fig. \ref{1dbench} as a function of the size of the corresponding three dimensional array.

\subsection{Three dimensional operators}\label{sec3doperators}
\begin{figure}
\begin{center}
\includegraphics[width=0.5\textwidth]{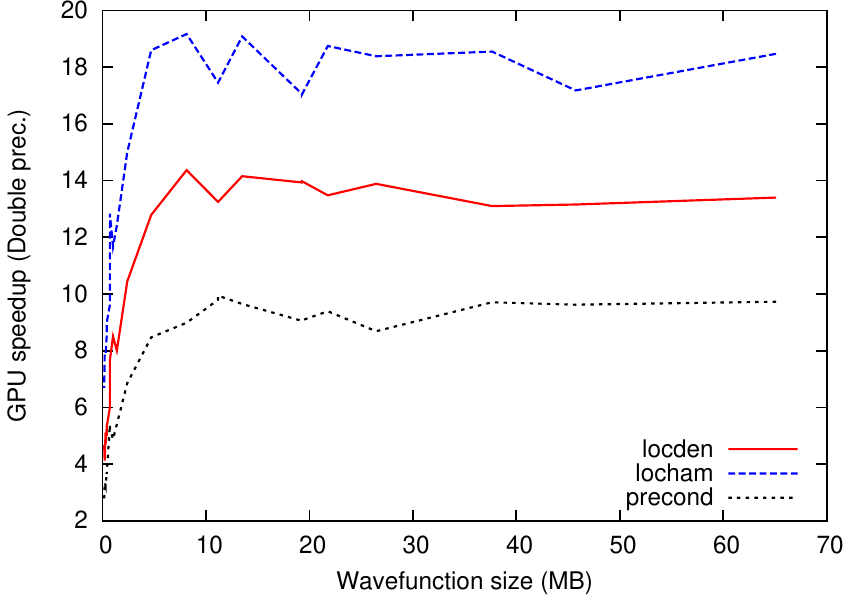}
\end{center}
\caption{Double precision speedup for the GPU version of the three-dimensional operators used in the BigDFT code as a function of the single wavefunction size.}
\label{3doperators}
\end{figure}

As described in the previous sections, to build a three-dimensional operation one must chain three times the corresponding one-dimensional GPU kernels.
We obtain in this way the three-dimensional wavelet transformations as well as the kinetic operator and the magic filter transformation (direct and transposed).
The multiplication with the potential and the calculation of the square of the wavefunction are performed via the application of some special GPU kernels, based on the same guidelines of the others.
Also the wavefunction compression-decompression can be calculated in the card.
The reduction operations, like for example the calculation of the potential and kinetic energy can be seen as linear algebra operations and can thus be performed with suitable calls to the corresponding \texttt{CUBLAS} routines.
The GPU speedup of the local density construction as well as the local hamiltonian application and of the preconditioning operation is represented in Fig. \ref{3doperators} as a function of the compressed wavefunction size.

\section{Toward a complete hybrid code}\label{memory}
In this section, we will discuss about the main issues for an efficient execution of hybrid code. 
The memory transfers between memories of CPU and GPU using the PCI-Express bus are well known to be potentially a bottleneck, and consequently a major obstacle to good performance.
So, we have to adapt the algorithm to reduce the memory transfers as most as possible. 
In our case, this can be done thanks to the scheduling of operations in the BigDFT code.  
Indeed, an optimized iteration of a single wavefunction is organized as follows:
\begin{itemize}
\item[I)] Density construction,
\item[II)] Poisson solver $\rightarrow$ Local potential ($V_H + V_\text{xc}$),
\item[III)] Local hamiltonian,
\item[IV)] Non-Local hamiltonian,
\item[V)] Wavefunction residue,
\item[VI)] Residue preconditioning,
\item[VII)] Wavefunction update,
\item[VIII)] Orthogonalization (Cholesky factorization).
\end{itemize}
The wavefunctions do not evolve between steps I) and III). 
So, they can be transferred on the GPU (global memory) before computing the density. 
After the step III), they can be sent back to the host memory (CPU), thus saving two memory transfers.
Morever, in this way, computation time is saved since the operations needed for the density calculation coincide with the first part of the operations for the local hamiltonian, see Sec. \ref{locden}.
Given the above scheduling of operations, the full CPU-GPU implementation of BigDFT code can be efficiently implemented.
In the current hybrid implementation, we can execute on the GPU the steps I), III) and VI) and also all BLAS routines performed in steps V) (DGEMM) and VIII) (DSYRK).
However, all other operations, such as LAPACK routines (step VIII) or the multiplication with  the nonlocal pseudopotentials (step IV) are still executed on the CPU and can be ported on GPU.
We left these implementation to future versions of the hybrid code.

\subsection{Parallel distribution}\label{paralleldist}
Hybrid architectures are becoming more and more popular with configurations of several CPU and GPU.  Typically, a configuration may be composed of two quad-cores processors (Intel Xeon or Nehalem) and two NVidia GPU. So, in this case, the two GPU have to be shared between the eight CPU cores. The problem of data distribution and load balancing is a major issue to achieve optimal performance. The operators implemented on GPU are parallelized within the orbital distribution scheme (see Sec. \ref{parallelcpu}).
This means that each core may host a subset of the orbitals, and apply the operators of Sec. \ref{sec3doperators} only to these wavefunctions.

A possible solution to the GPU sharing is to dedicate statically one GPU to one CPU core. 
So, in the common configuration, two CPU cores are more powerful because they have access to GPU. 
The six other CPU cores do not interact with the GPU.
Since the number of orbitals which may be assigned to each core can be adjusted, a possible way to handle the inhomogeneity CPU/GPU would be to assign more orbitals to the cores which have a GPU associated.
Such kind of approach can be realised thanks to the flexibility of the data distribution scheme of BigDFT.
However, it may be difficult for the end user to define an optimal repartition of the orbitals between the different cores for a generic system.

For this reason, we have designed an alternative approach where the GPU are completely shared by all CPU cores. 
The major consequence is that the set of orbitals is equally distributed to CPU cores.
Essentially, each of the CPU cores of a given node is allowed to use one of the GPU cards which are associated to the node, so that each card dialogues with a group of CPU cores.
Each GPU is associated to two semaphores, which control the memory transfers and the calculations.
In this way the memory transfers of the data between the different cores and the card can be executed asynchronously and overlapped with the calculations.
This is optimal since each orbital is processed independently.
The technical details of this implementation will be described elsewhere, via a more technical paper.
In the next section, we will study the performance of the hybrid BigDFT code.

\section{Performance evaluation of hybrid code}
The three-dimensional operator described in previous sections can be used in the full DFT code with the implementation of the memory transfers described in Sec.\ref{memory}.
As a test system, we used the ZnO crystal, which has a wurtzite bulk-like structure. Such system has a relatively high density of valence electrons so that the number of orbitals is rather large even for a moderate number of atoms.
Our calculations are performed on the hybrid part of the CINES IBLIS machine, which has 12 nodes, connected with an Infiniband 4X DDR connectX network, each node (2 Xeon X5472 quadri-core) connected with 2 GPU of a Tesla S1070 card.

We performed two kinds of tests. In the first one, to control the behaviour of the code for systems of increasing size, we performed a set of calculations for different supercells with increasing number of processors, such that the number of orbital per MPI process is kept constant. We performed a comparison for the same runs in which all the CPU cores have a GPU associated.
In the second test we kept fixed the size of the system and increased the number of MPI process such as to decrease the number of orbitals per core. 
We then controlled the speedup of each run by varying the number of cores associated to a GPU. This could be done thanks to the data transfer overlap which allows for multiple cores to share the same card, as described in Sec. \ref{paralleldist}.

\subsection{Homogeneous CPU/GPU repartition}
\begin{figure}
\begin{center}
\includegraphics[width=\textwidth]{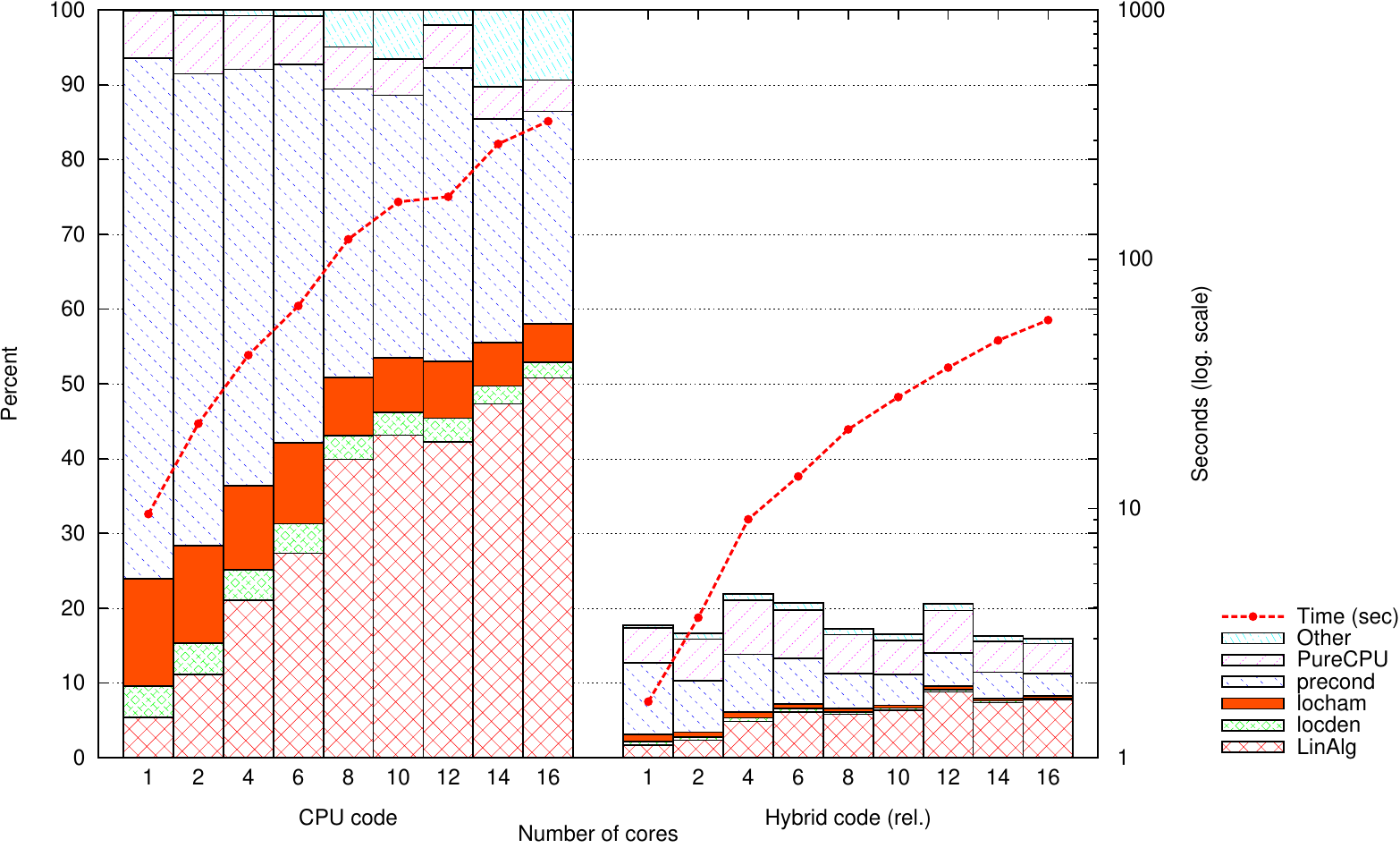}
\end{center}
\caption{Relative speedup of the hybrid DFT code wrt the equivalent pure CPU run, as a function of the number of orbitals. The calculation is performed in parallel such that each processor holds the same number of orbitals (36 in this figure). 
The number of atoms of each system is eight times the number of cores considered. Also the time in seconds for a single minimisation iteration is indicated, showing a speedup of a factor of around 6 with the hybrid CPU-GPU architecture, in double precision computation.}
\label{factor5}
\end{figure}

\begin{figure}
\begin{center}
\includegraphics[width=0.7\textwidth]{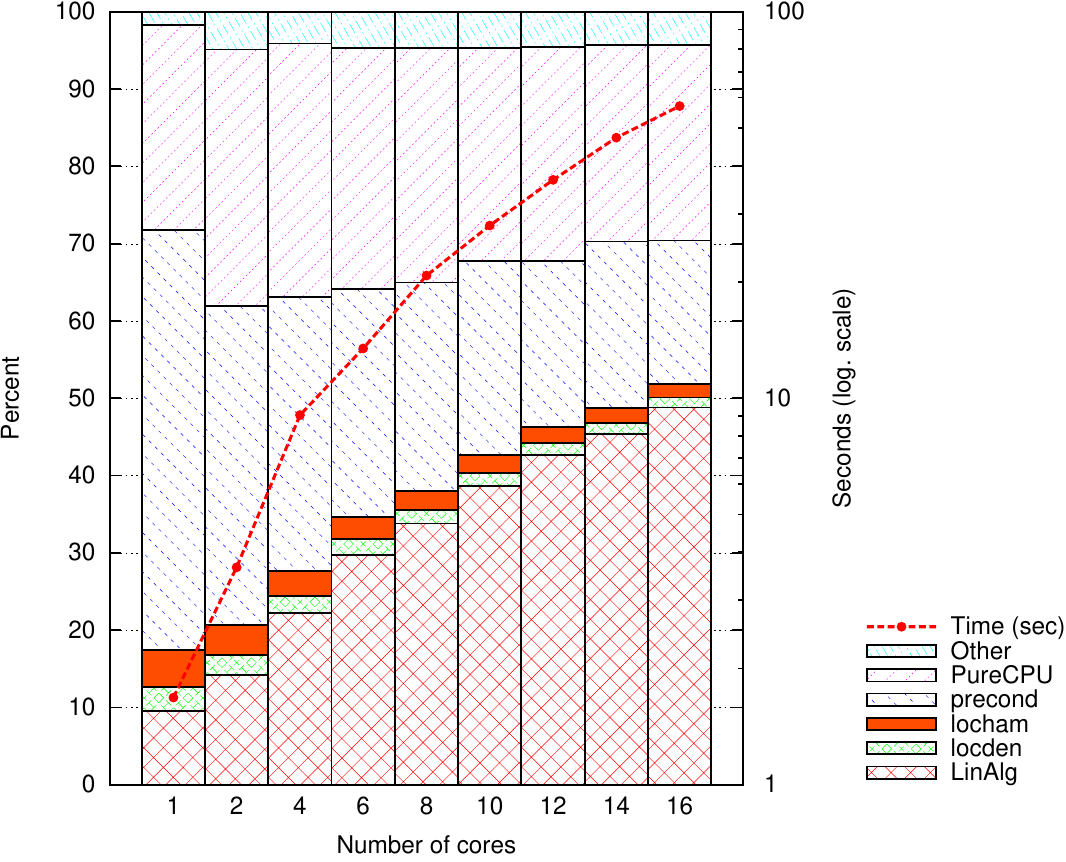}
\end{center}
\caption{Relative importance of different code section of the hybrid code, for the same systems analysed in Fig. \ref{factor5}.}
\label{pureGPU}
\end{figure}

\begin{figure}
\begin{center}
\includegraphics[width=0.7\textwidth]{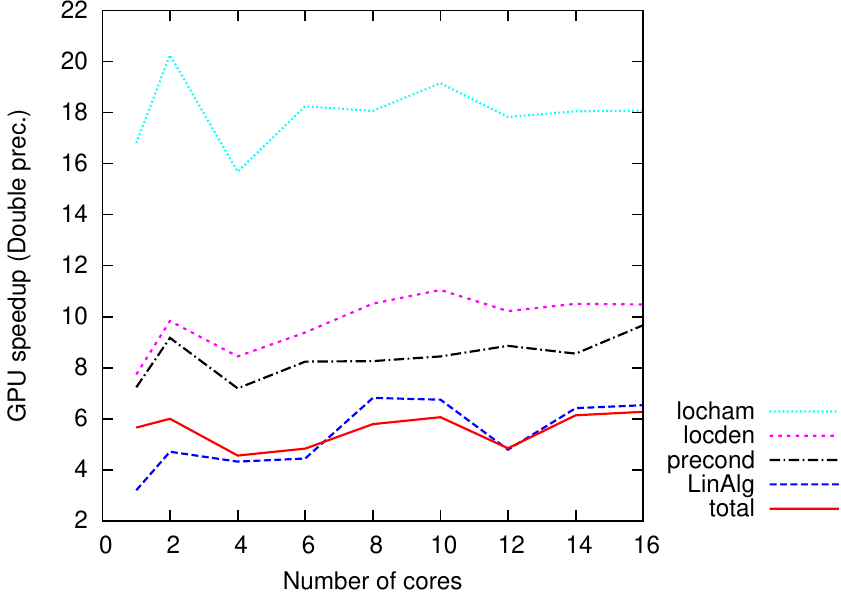}
\end{center}
\caption{Speedup of the GPU-related sections of the code for the systems of Fig.\ref{factor5}.}
\label{overallspeedup}
\end{figure}

For the first test, results are shown in Fig. \ref{factor5}. Our comparison shows an overall speedup of the whole code by a factor of around 6. 
This results is interesting and sounds very promising for a number of reasons.
First of all, as already discussed, not all the routines of the code were ported on GPU. 
We focus our efforts to the operators which can be written via a convolution. 
Also the application of the non-local part of the hamiltonian can be performed on the GPU, and we are planning to do this in further developments. 
Moreover, the actual implementation of the GPU convolutions can be further optimised. 
For example, in the preconditioner the wavelet transformation can be combined with the kinetic operator (as it is done in the CPU version) in order to define a faster GPU kernel. This whill allow to save more time, since the preconditioner still represents a considerable fraction of the overall time also in the GPU version, see Fig.\ref{pureGPU}.
Also the linear algebra operations can be further optimised. For the moment, only the calls to the BLAS routines were accelerated on the GPU, via suitable calls to the corresponding CUBLAS routines.
Also the LAPACK routines, which are needed to perform the orthogonalisation process, can be ported on GPU, with a considerable gain. 
Indeed the linear algebra operations represent the most expensive part of the code for very large systems (see \cite{BigDFT}). 
In figure \ref{overallspeedup}, the speedup of the different GPU-related parts of the runs of Fig.\ref{factor5} is shown.
As it can be seen, for large systems the overall speedup is dominated by the linear algebra operations. 
An optimisation of this section is then crucial for future improvements of the hybrid code.

\subsection{Inhomogeneous CPU/GPU repartition}
\begin{figure}
\begin{center}
\includegraphics[width=0.7\textwidth]{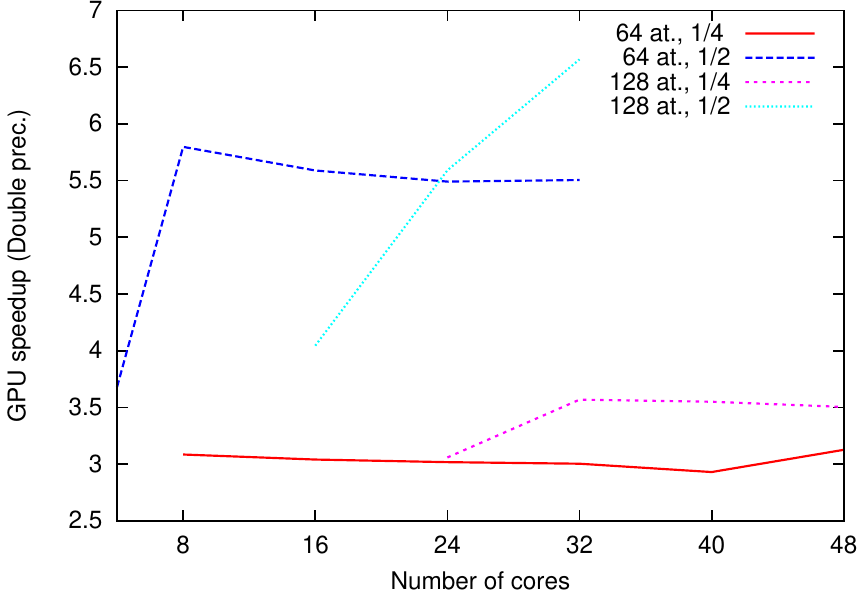}
\end{center}
\caption{Speedup of the full DFT code as a function of the number of cpu cores (i.e. MPI processes) for inhomogeneous repartitions of the number of GPU cards per core. The speedup is about 3.5 for a 1/4 repartition (four MPI processes per card), while for a 1/2 repartition it tends to be of the same order of the homogeneous case (see Fig.\ref{overallspeedup} for results on different systems).}
\label{hybrid}
\end{figure}

The second test mainly concerns the performances of the code in the case in which several MPI processes (and thus CPU cores) are associated to the same card.
We perform different runs for different repartitions of the card per core, such as to check the speedup by varying the ratio GPU/CPU on a hybrid run.
Results are plotted in Fig.\ref{hybrid}. 
We can see that the slowing down in the speedup with respect to the ratio GPU/CPU is less abrupt than it can be expected. 
In particular, results are not so much altered for a ratio of 50\% (1 GPU per 2 CPU cores) and only slow down of a factor of around two with a ratio of 25\%, i.e. four cores per card.
This is particularly encouraging since at the moment only the convolutions operators are de-syncronised by the semaphores (see Sec.\ref{paralleldist}), and the \texttt{BLAS} routines are executed at the same time on the card. Future improvements in this direction may allow us to better optimise the load on the cards such as to further increase the efficiency.

\section{Conclusions and perspectives}
The port of the principal sections of a electronic structure code over graphic processing units (GPU) has been shown.
Such GPU sections have been inserted in the complete code, in order to have a production DFT code which is able to run in a multi-GPU environment.
The DFT code we used, named BigDFT, is based on Daubechies wavelets, and has high systematic convergence properties, very good performances, and excellent efficiency on parallel computation.
The GPU implementation of the code we propose fully respects these properties. 
We use double precision calculations, and we may achieve considerable speedup for the converted routines (up to a factor of 20 for some operations). 
Our developments are fully compatible with the existing parallellisation of the code, and the communication between CPU and GPU does not affect the efficiency of the existing implementation.
The data transfers between the CPU and the GPU can be optimised in such a way to allow that more than one CPU core is associated to the same card.
This is optimal for modern hybrid supercomputer architectures in which the number of GPU cards is generally smaller than the number of CPU cores.
We test our implementation by running systems of variable number of atoms on a 12-node hybrid machine, with two GPU cards per node.
These developements produce an overall speedup on the whole code of a factor of around 6, also for a fully parallel run.
It should be stressed that, for these runs, our code has no hot-spot operations, and that all the sections which are ported on GPU contribute to the overall speedup.
Moreover, given the nature of the parallelisation of the BigDFT code, we expect these results to hold also on bigger systems in massive parallel hybrid environments, such as for example, the machine which is going to installed in France, at the CCRT, in the near future.

The GPU port of the routine was performed for fully periodic boundary conditions(BC), but also different BC can be implemented without altering the nature of the developments.
This is particularly interesting since it means that all these developments are totally compatible with the addition of new functionalities, like for example a linear scaling approach of the BigDFT code, which is under preparation.

Such results can however be further improved by optimising the present GPU routines, and by accelerating other sections of the code.
The hybrid BigDFT code, like its pure CPU counterpart, is available under GNU-GPL license and can be downloaded form the site in Ref.\cite{BigDFT}.
Up to our knowledge, it is the first time that a systematic electronic structure code has been able to run on hybrid (super)computers in (massively) parallel environement.
In summary, our results open a path toward the use of GPU for double precision density functional theory calculations and encourage us to proceed further in these direction.

\end{document}